\documentclass[prl,twocolumn,longbibliography,preprintnumbers,notitlepage,fleqn]{revtex4-2}

\usepackage{color,soul}

\usepackage{contour}
\usepackage{ulem}

\contourlength{0.8pt}

\newcommand{\myuline}[1]{%
  \uline{\phantom{#1}}%
  \llap{\contour{white}{#1}}%
}
\renewcommand{\emph}{\textit}

\usepackage[bookmarks = true, citecolor = blue, colorlinks = true, linkcolor = magenta, urlcolor = blue]{hyperref}

\usepackage[T1]{fontenc}			
\usepackage[sc,osf]{mathpazo}   	

\usepackage{amsmath}  			
\usepackage{amsfonts}  			
\usepackage{graphicx}   			
\usepackage{mathrsfs, amsthm, amssymb}
\usepackage{bbm, bm}

\usepackage{nicefrac}    			

\usepackage[braket, qm]{qcircuit}	

\usepackage{nicefrac} 

\usepackage{tikz}
\usetikzlibrary{calc,decorations.pathmorphing,shapes}

\newcounter{sarrow}

\newcommand{\indep}{\perp \!\!\! \perp}
\newcommand{\nonindep}{\not\!\perp\!\!\!\perp}

\begin{document}


\title{Causal reappraisal of the quantum three box paradox
}

\author{Pawel \surname{Blasiak}}
\email[E-mail: ]{pawel.blasiak@ifj.edu.pl}
\affiliation{Institute of Nuclear Physics Polish Academy of Sciences, PL-31342 Krak\'ow, Poland}

\author{Ewa \surname{Borsuk}}
\affiliation{Institute of Nuclear Physics Polish Academy of Sciences, PL-31342 Krak\'ow, Poland}

\begin{abstract}
Quantum three box paradox is a prototypical example of some bizarre predictions for intermediate measurements made on \textit{pre- and post-selected} systems. Although in principle those effects can be explained by \textit{measurement disturbance}, it is not clear what mechanisms are required to fully account for the observed correlations. In this paper, this paradox is scrutinised from the causal point of view. We consider an array of potential causal structures behind the experiment, eliminating those without enough explanatory power. This gives a means of differentiating between the various mechanisms in which measurement disturbance can propagate in the system. Specifically, we distinguish whether it is just the \textit{measurement outcome} or the \textit{full measurement context} that is required for the causal explanation of the observed statistics. We show that the latter is indispensable, but only when the full statistics is taken into account (which includes checking the third box too). Furthermore, we discuss the \textit{realism assumption} which posits the existence of preexisting values revealed by measurements. It is shown that in this case measurement disturbance is necessary. Interestingly, without the realism assumption, the original version of the paradox (with just two boxes considered for inspection) can be explained without resorting to any measurement disturbance. These various results illustrate the richness of the paradox which is better appreciated from the causal perspective.
\end{abstract}

\maketitle

A paradox builds upon a conflict between observed facts and preconceptions that we hold about them, with a view to elicit a revision of the latter for a deeper understanding of a given problem or phenomenon. On the one hand, this may be just a warning about a superficial understanding of the mathematics, which if correctly applied, does not lead to any contradictions. It is particularly true about the problems involving probability. On the other hand, a paradox can be an indication of a deeper misconception regarding the mechanisms at work, and thus potentially revealing something new about the nature itself. Therein lies the interest for the foundational issues of quantum theory and, in particular, the question about the causal nature behind its predictions. Notably, some remarkable results on quantum non-locality~\cite{Be93,BrCaPiScWe14,Sc19} or contextuality~\cite{KoSp67,ThKuLeSoKa16,BuCaGuKlLa21} are prime examples with a paradox and causality at the background.

A \textit{three box paradox}~\cite{AhVa91} is a flagship example of the \textit{pre- and post-selection} scenarios, in which some surprising predictions about intermediate measurements are made. It was originally proposed as an illustration of the so-called ABL rule~\cite{AhBeLe64} (after Aharonov, Bergmann and Lebowitz). For the three box paradox case  it makes a strange prediction regarding the position of a particle which is always found where it is looked for. This has sparked controversy regarding the nature of the paradox and conclusions that can be drawn from this bizarre effect~\cite{Ka99,Va99,Fi06,Ki03a,RaVa07,Ki07,Ma17}. The first objection concerns the presence of post-selection in the experiment, since the rejection of data is a potential source of non-causal correlations known as a \textit{selection bias}~\cite{Pe09}. The second problem stems from the possible role of \textit{measurement disturbance} in the experiment, since in this case the disturbance can propagate in the system making the information about the intermediate measurements available at the moment of post-selection. Those issues certainly affect interpretation of the paradox and thus need careful reassessment within a proper conceptual framework. 

In this work we tread the path eloquently expressed in Judea Pearl's~\cite{PeMa18} conjecture \textit{“that human intuition is organized around casual, not statistical, relations.”} It suggests that in an attempt at resolving a paradox one should rather focus on causal mechanisms behind the observed correlations. Not only this gives a way to the bottom of the paradox by explicating the implicit assumptions that we make, but sometimes may even offer something new about the causal mechanisms at work. Notably, the causal approach has recently gained a solid mathematical foundation in the works of Judea Pearl and others~\cite{Pe09,SpGlSc00,PeGlJe16} (which goes along similar lines as introduced by John Bell~\cite{Be93}). Some remarkable results in the field of causal inference pertinent to the present work include $d$\textit{-separation rules} and \textit{instrumental inequalities}. This novel approach has helped in resolving various conundrums in observational studies in epidemiology, computer science or social sciences~\cite{HeRo20,PeJaSc17,RuIm15}. Despite a fairly recent development, mostly outside of physics, those methods have already influenced the research in quantum foundations, see e.g.~\cite{WoSp15,ChKuBrGr15,RiAgVeJaSpRe15,RiGiChCoWhFe16,AlBaHoLeSp17,ChCaAgDiAoGiSc18,ChLePi18,BlPoYeGaBo21}.

In this article, we employ the tools of causal inference to analyse measurement disturbance and its impact on post-selection in the three box paradox. This approach allows differentiating between the various mechanisms in which measurement disturbance can propagate. We also bring to light some implicit assumptions about the realism that are typically made, which explains where the clash with our intuition might come from.

\vspace{0.15cm}

\textbf{\textsf{Three box paradox in a nutshell.}}---Consider an experiment with a system prepared at time $t_0$ in some state initial state $\rho_0$ on which at a later time $t_2$ a projective measurement $\mathcal{M}_2$ is made checking for state $\Pi^{\scriptscriptstyle post}$. Let us agree to retain only the positive results $M_2=1$, which are deemed a success, and discard all the rest $M_2=0$. This is the so-called \textit{pre- and post-selection scenario} (PPS). To make it more interesting we allow to make a measurement $\mathcal{M}_1^{\scriptscriptstyle C}$ in some intermediate time $t_1$ ($t_0<t_1<t_2$) described by a projection-valued measure (PVM) $\{\Pi_i^{\scriptscriptstyle C}\}$, where $C$ stands for the \myuline{\textit{choice}} of measurement in a given experimental run. Then, the conditional probability of obtaining outcome $M_1=i$ in the PPS scenario is given by the ABL rule~\cite{AhBeLe64}
\begin{eqnarray}\label{ABL}
P(M_1=i|M_2=1,C)&=&\tfrac{\text{Tr}\left[\Pi^{\scriptscriptstyle post}\right.\,\Pi^{\scriptscriptstyle C}_i\,\rho_0\,\left.\Pi^{\scriptscriptstyle C}_i\right]}{\sum_l\text{Tr}\left[\Pi^{\scriptscriptstyle post}\,\Pi^{\scriptscriptstyle C}_l\,\rho_0\,\Pi^{\scriptscriptstyle C}_l\right]}\,.
\end{eqnarray}
It is a straightforward application of Bayes' theorem to the joint probability
\begin{eqnarray}\label{joint}
P(M_1=i,M_2=j|C)&=&\text{Tr}\big[\Pi^{\scriptscriptstyle post}_j\,\Pi^{\scriptscriptstyle C}_i\,\rho_0\,\Pi^{\scriptscriptstyle C}_i\big]\,,
\end{eqnarray}
which obtains by the usual von Neumann-Lüders rule. Here the final measurement  $\mathcal{M}_2$ is described by the PVM $\{\Pi^{\scriptscriptstyle post}_j\}_{j\text{=}0,1}\equiv\{\mathbb{1}-\Pi^{\scriptscriptstyle post},\Pi^{\scriptscriptstyle post}\}$.

\begin{figure}[t]
\centering
\includegraphics[width=\columnwidth]{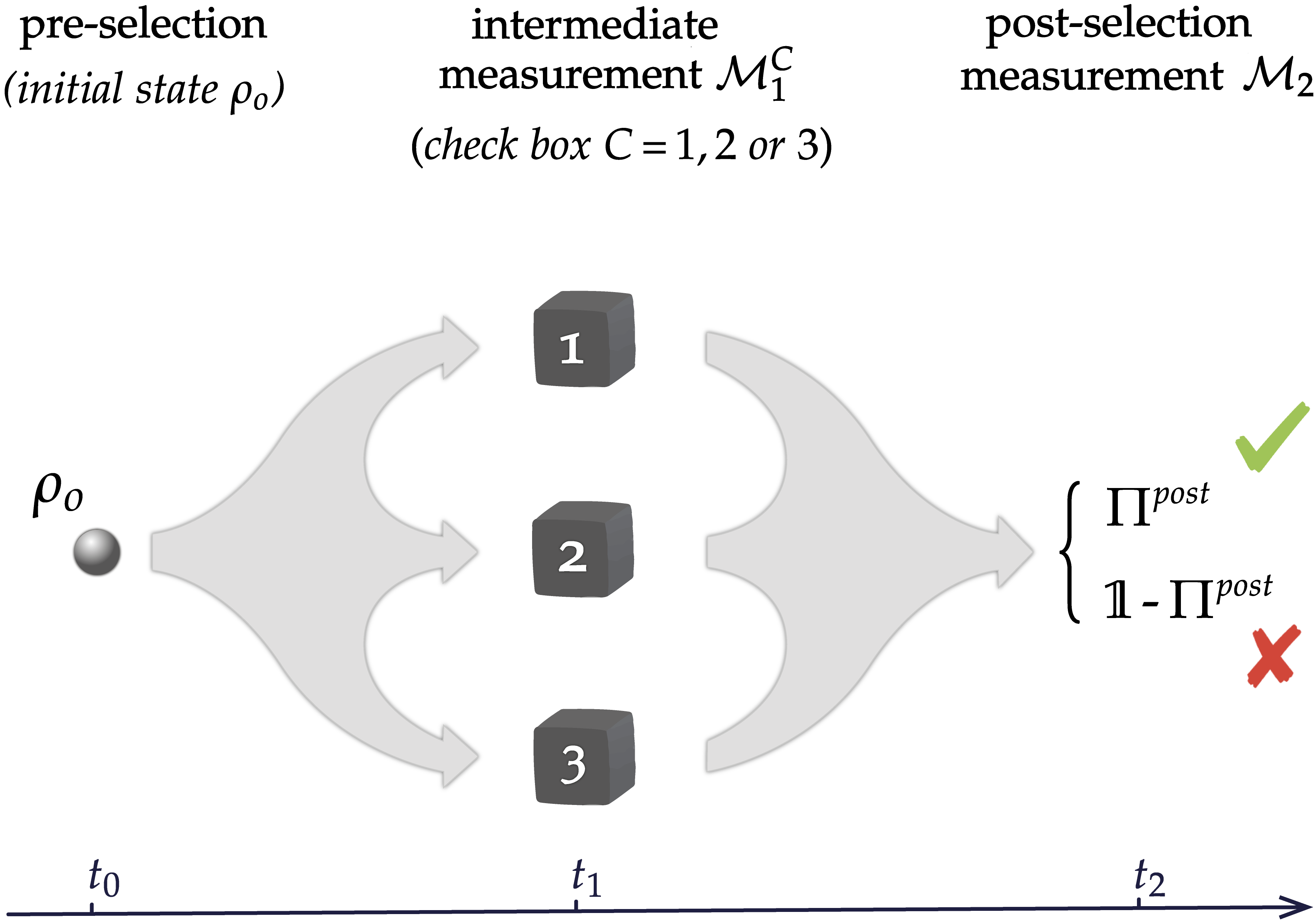}
\caption{\label{Fig-3-Box-Experiment}{\bf\textsf{\mbox{Three box experiment.}}} Consider a particle that can be in one of three boxes labelled $1$, $2$ and $3$. The system is \textit{pre-selected} in state $\rho_0$ and \textit{post-selected} in state $\Pi^{\scriptscriptstyle post}$. In each experimental trial we \textit{choose} a box $C=1,2,3$ checking whether the particle is there or not. Quantum mechanics makes a puzzling prediction that whichever box $C=1$ or $2$ we choose to look at, the particle will be always found there; see Eq.~(\ref{3-box}).} 
\end{figure}

A \textit{three box paradox}~\cite{AhVa91} is a specific realisation of the PPS scenario in which the intermediate measurements $\mathcal{M}_1^{\scriptscriptstyle C}$ are assigned deterministic conditional outcomes. It concerns a single particle that can be localised in one of three boxes labelled $1$, $2$ and $3$ described by the respective quantum states $\ket{1}$, $\ket{2}$ and $\ket{3}$. Suppose in the intermediate measurements we check whether the particle is in a given box $C=1,2,3$, or not, which is implemented by the PVM $\{\Pi^{\scriptscriptstyle C}_i\}_{i\text{=}0,1}\equiv\{\mathbb{1}-\op{C}{C},\op{C}{C}\}$. See Fig.~\ref{Fig-3-Box-Experiment}. Now, if we choose for the pre- and post-selected states, $\rho_0=\op{\phi}{\phi}$ and $\Pi^{\scriptscriptstyle post}=\op{\psi}{\psi}$, the following nonorthogonal pair
\begin{eqnarray}\label{pre-post-states}
\ket{\phi}=\tfrac{\ket{1}+\ket{2}+\ket{3}}{\sqrt{3}}&\quad\&\quad&\ket{\psi}=\tfrac{\ket{1}+\ket{2}-\ket{3}}{\sqrt{3}}\,,
\end{eqnarray}
then, from the ABL rule Eq.~(\ref{ABL}), we get
\begin{eqnarray}\label{3-box}
P(M_1=1|M_2=1,C)&=&1\qquad \text{for \ $C=1,2$}\,.
\end{eqnarray}
Hence a paradoxical conclusion that whatever box we check $C=1$ or $2$\,, the particle is always there. 

\begin{figure}[t]
\centering
\includegraphics[width=\columnwidth]{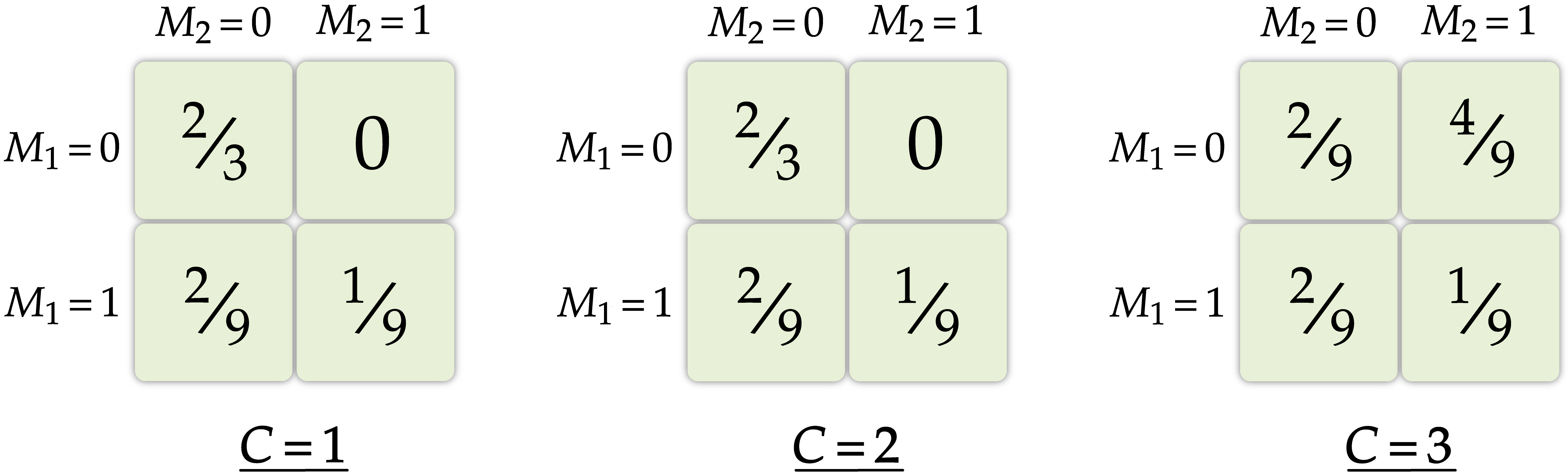}
\caption{\label{Fig-3-Box-Behaviour}{\bf\textsf{\mbox{Full statistics in three box experiment.}}} The joint probability $P(M_1,M_2|C)$ of obtaining measurement outcomes $M_1=0,1$ (i.e., particle not found or found) and $M_2=0,1$ (i.e., post-selection is a failure or success) for the \textit{choice} of experiment $C=1,2,3$ (i.e., which box to check).}
\end{figure}

The full statistics observed in the experiment is given in Fig.~\ref{Fig-3-Box-Behaviour}, which readily follows from Eq.~(\ref{joint}). This shows that for $C=1,2$ post-selection succeeds with the probability equal $P(M_2=1|C)=\nicefrac{1}{9}$~\footnote{We remark that this probability is the same as when no intermediate measurement $\mathcal{M}_1^{\scriptscriptstyle C}$ is made, in which case we also have $P(M_2=1|\,\text{without $\mathcal{M}_1^{\scriptscriptstyle C}$})=\text{Tr}\left[\Pi^{\scriptscriptstyle post}\,\rho_0\right]=|\!\ip{\psi\,}{\,\phi}\!|^2=\nicefrac{1}{9}$.}. Let us note in advance that although the original formulation of the paradox in Eq.~(\ref{3-box}) concerns just two (out of three possible) experimental choices $C=1,2$\,, it becomes more revealing and weird when we look at the full statistics $C=1,2,3$.

\vspace{0.15cm}

\textbf{\textsf{Causal picture of the three box experiment.}}---Let us consider the possible causal structures hiding behind the experiment which will be further assessed against their capacity for generating the three box statistics in Fig.~\ref{Fig-3-Box-Behaviour}. We need to decide about the variables deemed relevant for the description of the experiment and then ponder their causal relationships.
\vspace{0.15cm}

\mbox{\textsl{\textsf{\myuline{Pure causal setting}:}\ \vspace{0.1cm}}}\vspace{0.15cm}

In the description of the three box experiment there are three \textit{observed} variables
\begin{eqnarray}\nonumber
C&\!\!:& \text{choice of measurement setting ($C=1,2,3$)}\,,\\\nonumber
M_1&\!\!:& \text{outcome of measurement $\mathcal{M}_1^{\scriptscriptstyle C}$ ($M_1=0,1$)}\,,\\\nonumber
M_2&\!\!:& \text{outcome of measurement $\mathcal{M}_2$ ($M_2=0,1$)}\,.
\end{eqnarray}
Furthermore, let us postulate the existence of some \textit{unobserved} variable
\begin{eqnarray}\nonumber
\qquad\ \Lambda&\!\!:& \text{hidden (or latent) variable}\,.
\end{eqnarray}
This variable is aimed to describe any other factors relevant for the experiment (e.g., the details of preparation procedure). It is left unspecified in order not to restrict the range of possible explanations behind the observed correlations. We call it a \textit{pure causal} setting as it involves the least number of assumptions (this should be compared with the more restricted framework below).

\vspace{0.15cm}

\clearpage

\mbox{\textsl{\textsf{\myuline{Realist causal setting}:}\ \vspace{0.1cm}}}\vspace{0.15cm}

It is very instructive to consider in parallel the additional \textit{realism} assumption. This is a position  which derives from the worldview in which
physical objects and their properties exist, and the measurements reveal those preexisting values. In our case this view posits the existence of an additional variable
\begin{eqnarray}\nonumber
V&\!\!:& \text{position of the particle ($V=1,2,3$)}\,.
\end{eqnarray}
Since it is assumed to be revealed by the measurement $\mathcal{M}_1^{\scriptscriptstyle C}$, we have the following condition
\begin{eqnarray}\label{M1-real}
M_1(C,V)&:=&\delta_{\scriptscriptstyle C,V}\,.
\end{eqnarray}
Note that the variable $V$ is in principle \textit{unobserved}, but some information is revealed by the measurement $\mathcal{M}_1^{\scriptscriptstyle C}$ answering the question: "\textit{Is the particle in a given box $C$?}". We remark that those additional structural components make the \textit{realist causal} setting more restrictive compared to the \textit{pure causal} setting, as we shall see shortly.

\vspace{0.15cm}

\mbox{\textsl{\textsf{\myuline{Causal diagrams for the experiment}:}\ \vspace{0.1cm}}}\vspace{0.15cm}

\begin{figure}[t]
\centering
\includegraphics[width=1\columnwidth]{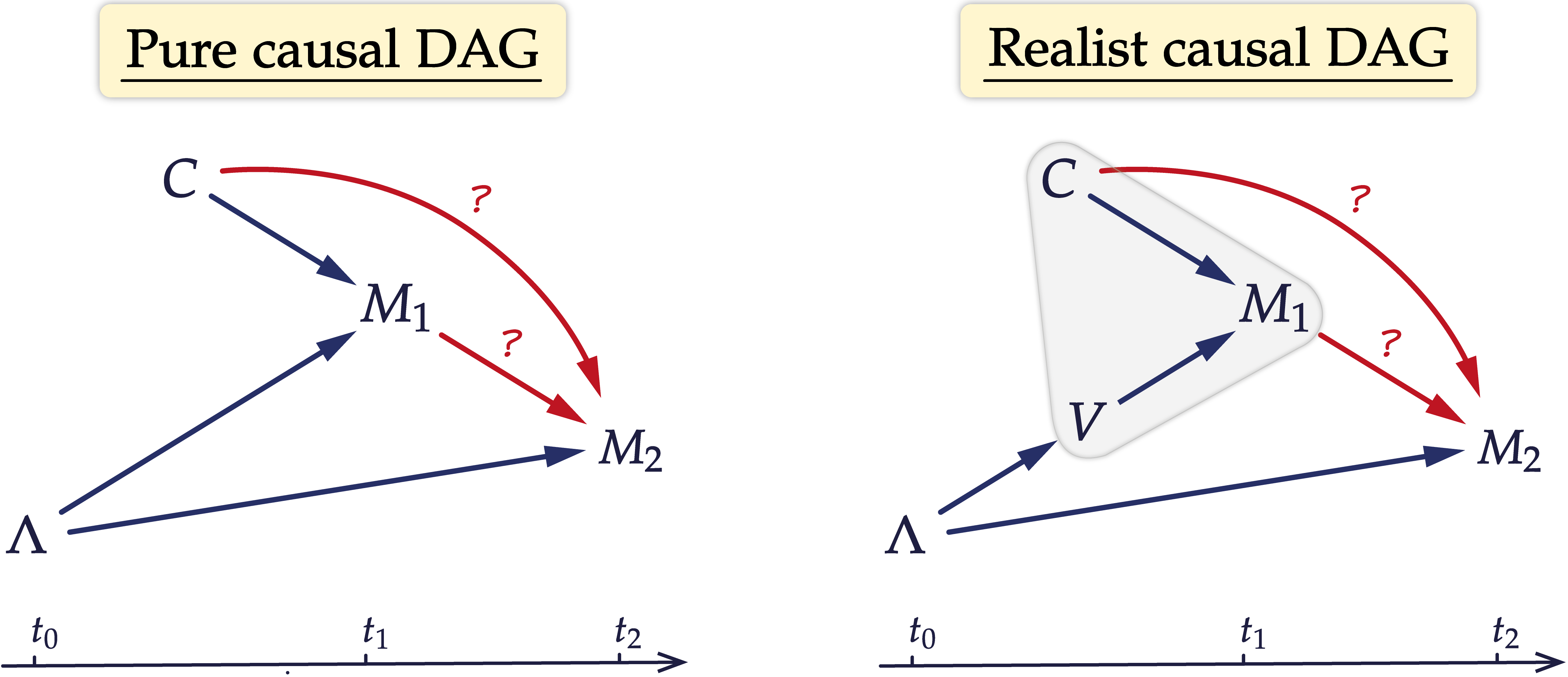}
\caption{\label{Fig-3-Box-Causal-DAGs}{\bf\textsf{\mbox{Causal diagrams for the three box experiment.}}} In both diagrams, the variables $C$, $M_1$ and $M_2$ are \textit{observed} in the experiment, whereas $\Lambda$ and $V$ are \textit{unobserved} (latent or hidden) variables. These are \textit{directed acyclic graphs} (DAGs) which are allowed by the temporal structure of the experiment (no retro-causation) and assuming $C$ to be a free variable. The diagram on the right includes an additional structure (shaded in grey) which reflects the realism assumption. In both diagrams the red arrows (with question marks) are responsible for different types of measurement disturbance propagating in the system, i.e. whether it is just the outcome $M_1$ or the full measurement context $C$ that affects $M_2$.
}
\end{figure}

Let us draw the diagrams compatible with the above two descriptions. In Fig.~\ref{Fig-3-Box-Causal-DAGs} the arrows represent cause-and-effect relationships between the variables. Observe that the temporal structure of the experiment allows to eliminate certain arrows in the diagrams. Namely, we assume only \textit{forward-in-time causation}. Also, since the choice of measurement is considered to be a \textit{free variable}, we assume there is no arrow incoming to $C$. [Note that although in principle in the realist causal DAG (on the right) there could be an arrow $V\rightarrow M_2$, it has not been drawn since we can always incorporate it in the arrow $\Lambda\rightarrow M_2$ (by appropriately modifying $\Lambda$).]

We note that the diagrams in Fig.~\ref{Fig-3-Box-Causal-DAGs} include \textit{all} arrows compatible with the experiment. In this paper we pose the question about the \textit{necessity} of arrows $M_1\rightarrow M_2$ and $C\rightarrow M_2$. Both are responsible for the causal effects of the intermediate measurement $\mathcal{M}_1^{\scriptscriptstyle C}$ in the experiment. The lack of both arrows means no measurement disturbance. Conversely, their presence is a sign of different types of disturbance propagating in the system:\vspace{0.3cm}

\textit{\parbox{0.95\columnwidth}{either it is just the measurement \myuline{outcome} $M_1$ or the full context specified by the \myuline{choice} of measurement (\myuline{parameter})  $C$ that affects the final measurement ${M}_2$.}}
\vspace{0.3cm}

\noindent Using the terminology borrowed from the analysis of Bell non-locality~\cite{Ja84} we call those arrows as
\begin{eqnarray}\nonumber
M_1\rightarrow M_2&:&\textit{outcome dependence,}\\\nonumber
C\rightarrow M_2&:&\textit{parameter dependence.}
\end{eqnarray}

Having specified causal structures of interest we may asses their potential for explaining the statistics of the three box paradox in Fig.~\ref{Fig-3-Box-Behaviour}.

\vspace{0.15cm}

\newpage

\textbf{\textsf{Main~results:~Inspection~of~causal~structures.}}---In~the causal inference field we are interested in verifying whether a given causal structure can explain the observed experimental behaviour, i.e., whether the statistics can be reproduced by some \textit{structural causal model} (SCM) consistent with a given causal DAG, cf.~\cite{Pe09,SpGlSc00,PeGlJe16,HeRo20,PeJaSc17,RuIm15,WoSp15,ChKuBrGr15,RiAgVeJaSpRe15,RiGiChCoWhFe16,AlBaHoLeSp17,ChCaAgDiAoGiSc18,ChLePi18,BlPoYeGaBo21}. Since adding arrows to the diagram extends its expressive power, we are looking for structures with the fewest number of arrows which can still explain the observed experimental behaviour.

As noted, the \textit{realist causal} setting  is more restrictive than the \textit{pure causal} setting (cf. Fig.~\ref{Fig-3-Box-Causal-DAGs}), i.e., not all behaviours compatible with the \textit{pure} causal DAG (on the left) are admitted by the \textit{realist} causal DAG (on the right). In the following we consider both cases separately. Furthermore, we also distinguish between the statistics in the three box experiment in Fig.~\ref{Fig-3-Box-Behaviour} for the \textit{full} choice of three measurements $C=1,2,3$, and the case \textit{limited} to the two choices $C=1,2$ (as in the original exposition of the paradox). This will make an interesting case regarding our perception of the paradox and its further ramifications as explained below.

\begin{figure*}
\centering
\includegraphics[width=0.94\textwidth]{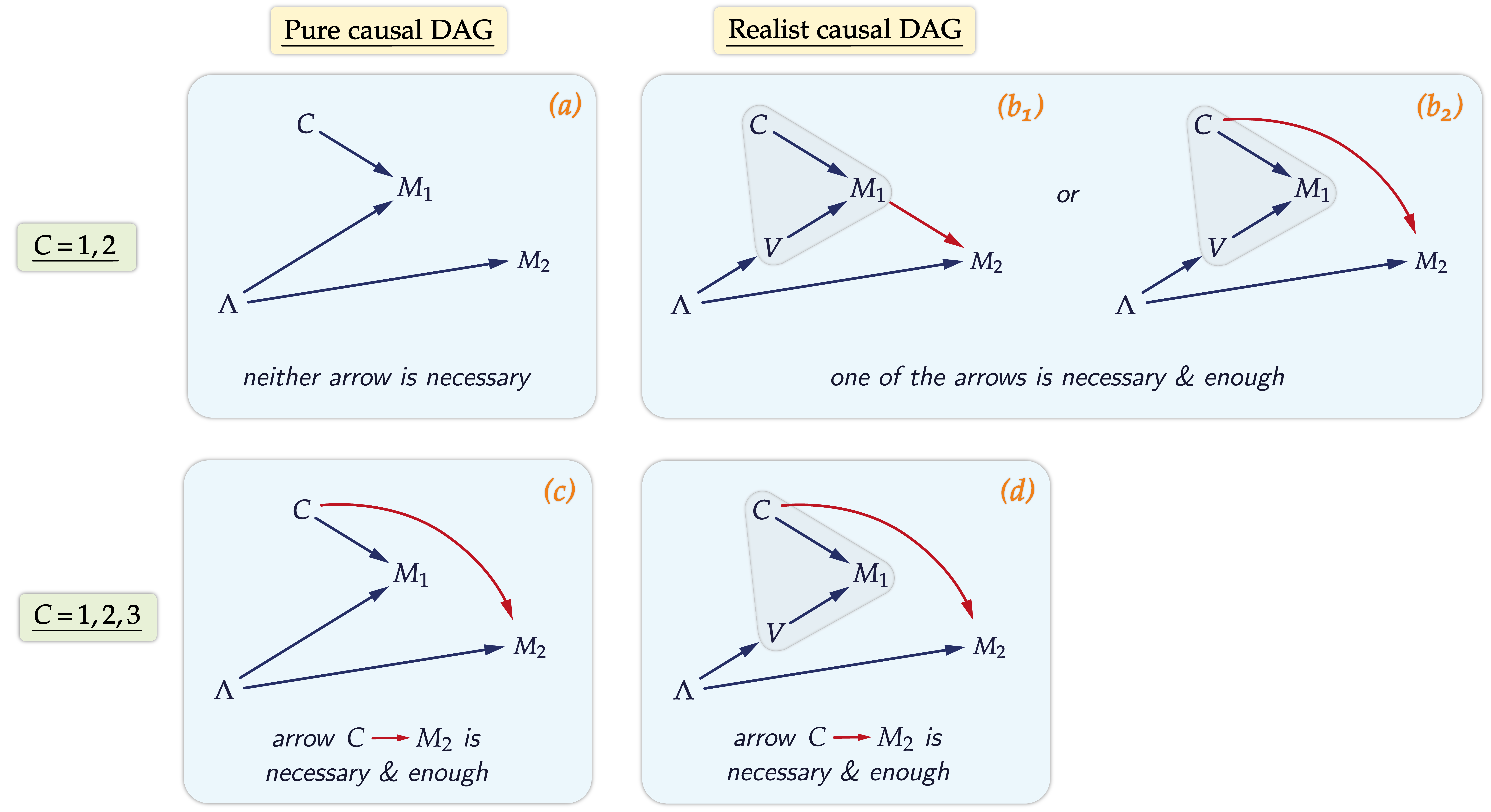}
\caption{\label{Fig-3-Box-Summary-DAGs}{\bf\textsf{\mbox{Summary of the results.}}} The table answers the question which red arrows in the two causal diagrams in Fig.~\ref{Fig-3-Box-Causal-DAGs} can be removed while still retaining their capacity for generating the statistics in the three box experiment in Fig.~\ref{Fig-3-Box-Behaviour}. There is a difference whether the full statistics is considered $C=1,2,3$ or just its part related to the 'paradoxical' choice of measurements $C=1,2$. This shows what kind of measurement disturbance (\textit{outcome} vs \textit{parameter dependence}) is required depending on the preferred worldview (\textit{pure} vs \textit{realist}) and the selection of measurements under consideration.}
\end{figure*}

Our results are summarised in Fig.~\ref{Fig-3-Box-Summary-DAGs}. For the proofs of necessity of the respective arrows see {\bf Appendix~A} (employing the tools of causal inference~\cite{Pe09}: $d$\textit{-separation criterion} and \textit{instrumental inequalities}). The proofs of sufficiency are given in {\bf Appendix~B} (this requires explicit construction of \textit{structural causal models} in each case). 

\vspace{0.15cm}

\textbf{\textsf{Conclusions.}}---Having analysed possible causal explanations of the three box experiment it is natural to ask why people find it surprising. Notably, the original formulation the paradox~\cite{AhVa91} concerns just two (out of three possible) experimental choices $C = 1, 2$\,; here the post-selected behaviour is deterministic which makes it  better suited for human judgement. In this case the paradox seems to arise from the tension between the assumption of realism and the deceptive impression, from how the paradox is phrased, as to the lack of measurement disturbance in the experiment. We showed that both assertions are intrinsically contradictory. The requirement of realism necessitates measurement disturbance of some sort (see $\textit{(b)}$ in Fig.~\ref{Fig-3-Box-Summary-DAGs}), i.e.,
\begin{eqnarray}\nonumber
\textsc{Realism}&\underset{\scriptscriptstyle C = 1, 2}{\Rightarrow}&\textsc{Measurement disturbance}\,.
\end{eqnarray}
Accordingly, if the disturbance is taken on board, then the paradoxical correlations in the post-selected regime can be explained as an instance of the \textit{selection bias}~\cite{Pe09} (see the proof of case $\textit{(b)}$ in {\bf Appendix~A}). 

Interestingly, in the \textit{pure causal} setting (\textit{no realism}) the paradox can be explained without measurement disturbance of any sort. In this case we showed, by constructing the explicit SCM, that confounding is just enough to explain the effect (see $\textit{(a)}$ in Fig.~\ref{Fig-3-Box-Summary-DAGs}), i.e.,
\begin{eqnarray}\nonumber
\textsc{No realism}&\underset{\scriptscriptstyle C = 1, 2}{/\!\!\!\!\!\!\Rightarrow}&\textsc{Measurement disturbance}\,.
\end{eqnarray}
As noted, both conclusions above concern the original formulation of the paradox with just two boxes being considered for inspection ($C=1,2$). 

Our discussion relies on a proper treatment of measurement disturbance as a genuine causal notion. This approach allows to make a further distinction between various types thereof, that is \textit{outcome} vs \textit{parameter dependence} represented by the respective arrows $M_1\rightarrow M_2$ and $C\rightarrow M_2$ in Fig.~\ref{Fig-3-Box-Causal-DAGs}. For this purpose the full statistics, which includes checking the third box ($C=1,2,3$) appears to be more interesting. It allows to show that \textit{parameter dependence} is actually necessary in both \textit{pure} and \textit{realist} framework (see $\textit{(c)}$ and $\textit{(d)}$ in Fig.~\ref{Fig-3-Box-Summary-DAGs}), i.e.,
\begin{eqnarray}\nonumber
\textsc{Full statistics}&\underset{\scriptscriptstyle C = 1, 2,3}{\Rightarrow}&\textsc{Parameter dependence}\,.
\end{eqnarray}
We also showed, by construction of the explicit SCM, that \textit{parameter dependence} is sufficient to explain the observed statistics (this can be also deduced from the general property regarding saturation of the model by the single arrow $C\rightarrow M_2$, as proved in Ref.~\cite{Ev16}).

Let us emphasise that a statement that an arrow is unnecessary does not mean that in reality it is not present (this only means that one can explain the statistics without its assistance). However, the necessity of an arrow implies that it cannot be replaced by any other arrow and still correctly reproduce the statistics. In this sense the causal DAGs in Fig.~\ref{Fig-3-Box-Summary-DAGs} are the minimal structures compatible with the observed statistics under the given \textit{realist}/\textit{pure} assumption.

\vspace{0.15cm}

\textbf{\textsf{Discussion.}}---It should be remarked that for the three box paradox there is no physical principle that would prohibit the propagation of measurement disturbance in the experiment (like the locality principle for Bell experiments~\cite{Be93,BrCaPiScWe14,Sc19}). This is because thePPS paradigm consists of a sequence of two measurements $\mathcal{M}_1$ and $\mathcal{M}_1^{\scriptscriptstyle C}$, with the second one $\mathcal{M}_2$ being made on the whole system (i.e., even if the boxes can be kept separate while checking a given box $C$ in the measurement $\mathcal{M}_1^{\scriptscriptstyle C}$, all of them have to merge together to implement the measurement $\mathcal{M}_2$). This makes all the information about the outcome $\mathcal{M}_1^{\scriptscriptstyle C}$ and the parameter $C$ in principle available upon post-selection (i.e., neither of the disturbance arrows can be excluded by some fundamental principle). It is also clearly seen in the experimental realisations of the three box paradox~\cite{ReLuSt04,GeRoMaBlBeMaTw13}. We also note the existence of a generic local hidden variable model of a single particle in arbitrary linear optical circuits (where the measurement disturbance propagates locally too)~\cite{Bl18}. 

In this paper we focused on the assessment of the role of the various causal mechanisms capable of generating the statistics observed in the quantum three box paradox. Such an approach seems to be more revealing regarding the structure of measurement disturbance than the mere acknowledgement of its presence of some sort. The analysis based on the \textit{instrumental inequalities} shows the necessity of the \textit{parameter dependence} in the system (but interestingly, only when the full statistics is taken into account). Furthermore, the use of the $d$\textit{-separation criterion} allows to see the paradox as a case of the \textit{selection bias}~\cite{Pe09} (cf. Ref.~\cite{FeCo14,Va14} for a related issue of post-selection in the interpretation of weak values~\cite{AhAlVa88}). We remark that in our discussion we take the conservative point of view with single measurement outcomes (cf. the many-worlds interpretation) as well as exclude backward-in-time causation. 

Let us note that the PPS paradoxes are can be turned into the proofs of contextuality~\cite{LeSp05a,LeSp05,Pu14,PuLe15}. Since in principle all those effects can be attributed to measurement disturbance, it is natural to ask about the possible causal mechanisms allowing for this to be fully attained. This makes the question about the causal resources (here taken as different sorts of arrows) that are enough to explain a given class of contextual effects is an interesting research problem.
This work provides a strong hint that \textit{parameter dependence} is in general indispensable. However, we have also seen that it may be superfluous in some restricted settings (cf. $C=1,2$ vs $C=1,2,3$ in Fig.~\ref{Fig-3-Box-Summary-DAGs}). We note in passing an interesting question regarding the role of signalling in our argument. Observe that the marginals of $M_2$ in the behaviour in Fig.~\ref{Fig-3-Box-Behaviour} change only when $C=3$ is also considered, and it is where the \textit{parameter dependence} in the three box statistics can be proved (this should be compared with the non-signalling PR boxes for which the \textit{outcome dependence} is just enough). We leave it as a curious remark regarding contextuality in the presence of signalling~\cite{DzKuLa15,KuDzLa15}.

Finally, we mention the existence of a several related PPS paradoxes discussed in the literature, e.g.~\cite{AhVa03,AhVa08,Va13,AhPoRoSk13,AhCoPoSaStTo16,AhCoLaEl17}. Their resemblance to the three box paradox suggests that the causal approach will shed more light there too.

\vspace{0.15cm}

\textbf{\textsf{Acknowledgments.}}---We acknowledge helpful discussions with Jarek~Duda, Marcin~Markiewicz and Rafal~Staszewski. We also thank Elie~Wolfe for bringing Ref.~\cite{Ev16} to our attention and comments on the overall framework of this work. 

\bibliography{CombQuant}

\newpage

\appendix

\section{Appendix A:\ \ {Proofs of necessity}}\label{Appendix-Necessity}\vspace{-0.1cm}

Here we answer the question which of the two red arrows in the causal DAGs in Fig.~\ref{Fig-3-Box-Causal-DAGs} are necessary in order to reproduce the statistics in Fig.~\ref{Fig-3-Box-Behaviour}. For the proofs we use the modern tools of causal inference~\cite{Pe09}, i.e. $d$\textit{-separation criterion} (case \textit{(b)} in Fig.~\ref{Fig-3-Box-Summary-DAGs}) and \textit{instrumental inequalities} (case \textit{(c)} \& \textit{(d)} in Fig.~\ref{Fig-3-Box-Summary-DAGs}).

\begin{proof}[\myuline{Proof of case \textit{(b)} in Fig.~\ref{Fig-3-Box-Summary-DAGs}}]\ \vspace{0.1cm}\\
Let us consider the realist causal framework with just two measurement choices $C=1,2$. In that case, from the Eq.~(\ref{3-box}), we have $P(M_1=1|M_2=1,C)=1$. As a consequence of the assumption Eq.~(\ref{M1-real}) it means that $V=C$ whenever $M_2=1$. This necessitates the following conditional dependence $V\nonindep C\,|\,M_2$. However, in the realist causal DAG with both arrows $M_1\rightarrow M_2$ and $C\rightarrow M_2$ absent (cf. Fig.~\ref{Fig-3-Box-Causal-DAGs} on the right), the variables $C$ and $V$ are $d$\textit{-separated} conditioned on $M_2$ (since the only path $C\rightarrow M_1\leftarrow V$ is blocked by the collider $M_1$) which entails their statistical independence $V\indep C\,|\,M_2$. This contradiction means that at least one of those arrows \myuline{must} be present in the realist causal DAG. Indeed, either arrow $M_1\rightarrow M_2$ or $C\rightarrow M_2$ lifts the $d$\textit{-separation} by opening the respective path $V\leftarrow\Lambda \rightarrow M_2\leftarrow M_1\leftarrow C$ or $V\leftarrow\Lambda\rightarrow M_2\leftarrow C$ (here $M_2$ is not a collider because of conditioning). This opens a way for both variables become correlated (whether it is enough with a single arrow is proved by an explicit SCM in the {\bf Appendix~B}). In the field of causal inference the phenomenon of correlation due to conditioning is known as a \textit{selection bias} or \textit{Berkson’s paradox}~\cite{Pe09}. 
\end{proof}

\begin{proof}[\myuline{Proof of case \textit{(c)} \& \textit{(d)} in Fig.~\ref{Fig-3-Box-Summary-DAGs}}]\ \vspace{0.1cm}\\
Now we are concerned with the full choice of measurements $C=1,2,3$. For the prove it is enough to consider the \textit{pure} causal DAG, i.e. case \textit{(c)} in Fig.~\ref{Fig-3-Box-Summary-DAGs} (since this framework is more permissive, the necessity for \textit{(c)} automatically carries over to the \textit{realist} case \textit{(d)}). 

Let us reformulate our problem in the causal inference terms. Suppose we admit the arrow $M_1\rightarrow M_2$ in the \textit{pure} causal DAG (see Fig.~\ref{Fig-3-Box-Causal-DAGs} on the left) and ask about the necessity of the arrow $C\rightarrow M_2$. [Note that proving the necessity of arrow $C\rightarrow M_2$ in this case will entail its necessity when the arrow $M_1\rightarrow M_2$ is missing, since adding the latter is only increases the expressive power of the DAG).] This can be recast as the \textit{instrumental scenario}~\cite{Pe95b,Pe09}, where the instrument $C$ is used for determining causal influence of  $M_1$ on $M_2$, both of which are affected by some unobserved (or latent) variable $\Lambda$. The crucial assumption in this scenario is the absence of any arrow (in/out)going from $C$ except for $C\rightarrow M_1$. It appears that this assumption can be tested by the so called \textit{instrumental inequalities}~\cite{Pe95b,Pe09} which adopted to our case take the form
\begin{eqnarray}\label{Instrument-Ineq-3-Box}
\max_{i}\,\sum_{j}\,\big[\max_{k}P(M_1\!=\!i,M_2\!=\!j|C\!=\!k)\big]&\leqslant&1\,,
\end{eqnarray}
where $i,j=0,1$ and $k=1,2,3$. Violation of those inequalities by the observed statistics testifies to the presence of some other arrow (in/out)going from $C$ (whatever the character of the arrow $M_1\rightarrow M_2$). In our case this could be only the arrow $C\rightarrow M_2$, and hence a way to check  its necessity.

To unfold the condition Eq.~(\ref{Instrument-Ineq-3-Box}) we observe that it is equivalent to the following set of equations
\begin{eqnarray}\nonumber
P(M_1\!=\!0,M_2\!=\!0|C\!=\!k)+P(M_1\!=\!0,M_2\!=\!1|C\!=\!l)&\leqslant&1
\\\nonumber
P(M_1\!=\!1,M_2\!=\!0|C\!=\!k)+P(M_1\!=\!1,M_2\!=\!1|C\!=\!l)&\leqslant&1\\\nonumber
P(M_1\!=\!0,M_2\!=\!1|C\!=\!k)+P(M_1\!=\!0,M_2\!=\!0|C\!=\!l)&\leqslant&1\\\nonumber
P(M_1\!=\!1,M_2\!=\!1|C\!=\!k)+P(M_1\!=\!1,M_2\!=\!0|C\!=\!l)&\leqslant&1
\\\label{Instrument-Ineq-3-Box-explicit}
\end{eqnarray}
where $kl=12,13,23$. Now it is straightforward to check that the statistics in Fig.~\ref{Fig-3-Box-Behaviour} violates those inequalities for $kl=13$ and $23$. For example, for $kl=23$ in the first line in Eq.~(\ref{Instrument-Ineq-3-Box-explicit}) we get
\begin{eqnarray}
\nicefrac{2}{3}\ +\ \nicefrac{4}{9}\ =\ \nicefrac{10}{9}\ \ {>}\ \ 1\,.
\end{eqnarray}
Therefore, if the full choice of measurements $C=1,2,3$ is considered, then the arrow $C\rightarrow M_2$ \myuline{must} be present in the causal DAG in Fig.~\ref{Fig-3-Box-Causal-DAGs} (either \textit{pure} or \textit{realist}), if it is to reproduce the statistics in Fig.~\ref{Fig-3-Box-Behaviour} (this single arrow is also enough as proved by an explicit SCM in the {\bf Appendix~B}). We note that for the limited choice $C=1,2$ (i.e., $kl=12$) the instrumental inequalities Eqs.~(\ref{Instrument-Ineq-3-Box-explicit}) remain inviolate, which means that in that case the arrow is unnecessary (in full agreement with the case \textit{(a)} and \textit{($\textit{b}_\textit{1}$)} in Fig.~\ref{Fig-3-Box-Summary-DAGs}).
\end{proof}
\vspace{-0.4cm}

\section{Appendix B:\ \ Proofs of sufficiency}\label{Appendix-Necessity}\vspace{-0.1cm}

Here we give a repository of explicit \textit{structural causal models} (SCMs) proving the sufficiency of the considered causal DAGs for each case in Fig.~\ref{Fig-3-Box-Summary-DAGs}. In the following we just specify the structural equations for the respective SCMs and observe that the joint probability distributions $P(M_1\!=\!i,M_2\!=\!j|C\!=\!k)$ obtain by a straightforward application of the product decomposition (Markov property) for the associated causal DAGs. 

\begin{proof}[\myuline{Proof of case \textit{(a)} in Fig.~\ref{Fig-3-Box-Summary-DAGs}}]\ \vspace{0.1cm}\\
Note that the joint probability distribution generated by the causal DAG $\textit{(a)}$ in Fig.~\ref{Fig-3-Box-Summary-DAGs} has the following decomposition
\begin{eqnarray}
&&\!\!\!\!\!\!\!\!\!\!\!\!\!\!\!\!P(M_1\!=\!i,M_2\!=\!j|C\!=\!k)\nonumber\\
&&\!\!\!\!\!\!\!\!\!\!\!\!=\ \sum_{\lambda}\ P(M_1\!=\!i|C\!=\!k,\Lambda\!=\!\lambda)\cdot P(M_2\!=\!j|\lambda)\cdot P(\Lambda\!=\!\lambda)\,.\nonumber\\\label{Joint-Distr-A}
\end{eqnarray}

Now, in order to recover the statistics in Fig.~\ref{Fig-3-Box-Behaviour} for $C=1,2$  it is enough to consider a Bernoulli distributed hidden variable $\Lambda\sim\text{Ber}\,(\nicefrac{1}{3})$, i.e. we have $\Lambda=0,1$ and 
\begin{eqnarray}
P(\Lambda=0)=\nicefrac{2}{3}\quad\&\quad P(\Lambda=1)=\nicefrac{1}{3}\,.
\end{eqnarray}
Then, we put the following structural equations compatible with the diagram in Fig.~\ref{Fig-3-Box-Summary-DAGs}\,\textit{(a)}
\begin{eqnarray}
M_1(C,\Lambda)&:=&\Lambda\,,\\
M_2(\Lambda,N)&:=&\Lambda\cdot N\,,
\end{eqnarray}
where $N\sim\text{Ber}\,(\nicefrac{1}{3})$ is an independent  noise variable having a Bernoulli distribution. This defines the SCM, which via Eq.~(\ref{Joint-Distr-A}), gives the correct statistics in Fig.~\ref{Fig-3-Box-Behaviour} for $C=1,2$ (only!).
\end{proof}

\begin{proof}[\myuline{Proof of case \textit{(b)} in Fig.~\ref{Fig-3-Box-Summary-DAGs}}]\ \vspace{0.1cm}\\
Let us start by justifying the sufficiency of the causal DAG \textit{($\textit{b}_\textit{1}$)} in Fig.~\ref{Fig-3-Box-Summary-DAGs}. In this case, the joint probability distribution has the following decomposition
\begin{eqnarray}
&&\!\!\!\!\!\!\!\!\!\!\!\!\!\!\!\!P(M_1\!=\!i,M_2\!=\!j|C\!=\!k)\nonumber\\
&&\!\!\!\!\!\!\!\!\!\!\!\!=\ \sum_{\lambda,v}\ P(M_1\!=\!i|C\!=\!k,V\!=\!v)\cdot P(M_2\!=\!j|M_1\!=\!i,\Lambda\!=\!\lambda)\nonumber\\
&&\ \ \ \ \cdot\ P(V\!=\!v|\Lambda\!=\!\lambda)\cdot P(\Lambda\!=\!\lambda)\,.
\label{Joint-Distr-B1}
\end{eqnarray}

In this case we posit a uniformly distributed hidden variable $\Lambda\sim\text{Uni}\,(1,3)$, i.e. we have $\Lambda=1,2,3$ with
\begin{eqnarray}
P(\Lambda=i)=\nicefrac{1}{3}\qquad \text{for}\ \ i=1,2,3\,.
\end{eqnarray}
The following structural equations define an SCM compatible with the diagram in Fig.~\ref{Fig-3-Box-Summary-DAGs}\,\textit{($\textit{b}_\textit{1}$)}
\begin{eqnarray}
M_1(C,V)&:=&\delta_{\scriptscriptstyle C,V}\,,\qquad\qquad\text{[cf. Eq.~(\ref{M1-real})]}\\
V(\Lambda)&:=&\Lambda\,,\\
M_2(M_1,\Lambda)&:=&M_1\cdot N\,,
\end{eqnarray}
where $N\sim\text{Ber}\,(\nicefrac{1}{3})$ is a noise variable with a Bernoulli distribution. This is enough to recover, via Eq.~(\ref{Joint-Distr-B1}), the statistics in Fig.~\ref{Fig-3-Box-Behaviour} for $C=1,2$ (only!).

The sufficiency of the DAG \textit{($\textit{b}_\textit{2}$)} in Fig.~\ref{Fig-3-Box-Summary-DAGs} follows immediately from the model in the case \textit{(c)} below which works for all $C=1,2,3$. 
\end{proof}

\vspace{0.5cm}

\begin{proof}[\myuline{Proof of case \textit{(c)} in Fig.~\ref{Fig-3-Box-Summary-DAGs}}]\ \vspace{0.1cm}\\
Here the  joint probability distribution generated by the causal DAG \textit{(c)} in Fig.~\ref{Fig-3-Box-Summary-DAGs} has the following decomposition
\begin{eqnarray}
&&\!\!\!\!\!\!\!\!\!\!\!\!\!\!\!\!P(M_1\!=\!i,M_2\!=\!j|C\!=\!k)\nonumber\\
&&\!\!\!\!\!\!\!\!\!\!\!\!=\ \sum_{\lambda}\ P(M_1\!=\!i|C\!=\!k,\Lambda\!=\!\lambda)\nonumber\\
&&\ \ \ \ \cdot\ P(M_2\!=\!j|C\!=\!k,\Lambda\!=\!\lambda)\cdot P(\Lambda\!=\!\lambda)\,.
\label{Joint-Distr-C}
\end{eqnarray}

We need an SCM which reconstructs the statistics in Fig.~\ref{Fig-3-Box-Behaviour} for $C=1,2,3$ (all of them!). Consider a uniformly distributed hidden variable $\Lambda\sim\text{Uni}\,(1,3)$ which takes three values $\Lambda=1,2,3$. Then we define the following set of structural equations in accord with the diagram in Fig.~\ref{Fig-3-Box-Summary-DAGs}\,\textit{(c)}
\begin{eqnarray}\label{C1}
M_1(C,\Lambda)&:=&\delta_{\scriptscriptstyle C,\Lambda}\,,\qquad\qquad\qquad\text{[cf. Eq.~(\ref{M1-real})]}\\\label{C2}
M_2(C,\Lambda)&:=&\\\nonumber
&&\!\!\!\!\!\!\!\!\!\!\!\!\!\!\!\!\!\!\!\left\{
\begin{array}{ll}
\delta_{\scriptscriptstyle C,\Lambda}\cdot N&\ \text{for}\ C=1,2\,,\vspace{0.15cm}\\
(1-\delta_{\scriptscriptstyle C,\Lambda})\cdot (1-N)+\delta_{\scriptscriptstyle C,\Lambda}\cdot N&\ \text{for}\ C=3\,,
\end{array}
\right.
\end{eqnarray}
where $N\sim\text{Ber}\,(\nicefrac{1}{3})$ is a noise variable with a Bernoulli distribution. Such a definition of the SCM recovers, via Eq.~(\ref{Joint-Distr-C}), the full statistics in Fig.~\ref{Fig-3-Box-Behaviour} for all $C=1,2,3$.
\end{proof}

\begin{proof}[\myuline{Proof of case \textit{(d)} in Fig.~\ref{Fig-3-Box-Summary-DAGs}}]\ \vspace{0.1cm}\\
This case is a straightforward extension of the SCM in \textit{(c)}. Here the causal DAG \textit{(d)} in Fig.~\ref{Fig-3-Box-Summary-DAGs} entails decomposition of the joint probability distribution in the form
\begin{eqnarray}
&&\!\!\!\!\!\!\!\!\!\!\!\!\!\!\!\!P(M_1\!=\!i,M_2\!=\!j|C\!=\!k)\nonumber\\
&&\!\!\!\!\!\!\!\!\!\!\!\!=\ \sum_{\lambda,v}\ P(M_1\!=\!i|C\!=\!k,V\!=\!v)\cdot P(M_2\!=\!j|C\!=\!k,\Lambda\!=\!\lambda)\nonumber\\
&&\ \ \ \ \cdot\ P(V\!=\!v|\Lambda\!=\!\lambda)\cdot P(\Lambda\!=\!\lambda)\,.
\label{Joint-Distr-D}
\end{eqnarray}

Following the constriction in case \textit{(c)} we take the hidden variable $\Lambda=1,2,3$ with a uniform distribution $\Lambda\sim\text{Uni}\,(1,3)$. Then we introduce an additional variable $V=1,2,3$ and equate it with $\Lambda$. This trivial extension provides the required SCM which takes the following explicit form (cf. Eqs.~(\ref{C1})-(\ref{C2}))
\begin{eqnarray}
M_1(C,V)&:=&\delta_{\scriptscriptstyle C,V}\,,\qquad\qquad\qquad\text{[cf. Eq.~(\ref{M1-real})]}\\
V(\Lambda)&:=&\Lambda\,,\\
M_2(C,\Lambda)&:=&\\\nonumber
&&\!\!\!\!\!\!\!\!\!\!\!\!\!\!\!\!\!\!\!\left\{
\begin{array}{ll}
\delta_{\scriptscriptstyle C,\Lambda}\cdot N&\ \text{for}\ C=1,2\,,\vspace{0.15cm}\\
(1-\delta_{\scriptscriptstyle C,\Lambda})\cdot (1-N)+\delta_{\scriptscriptstyle C,\Lambda}\cdot N&\ \text{for}\ C=3\,,
\end{array}
\right.
\end{eqnarray}
where $N\sim\text{Ber}\,(\nicefrac{1}{3})$ is again a Bernoulli noise variable.
Those structural equations comply with the diagram in Fig.~\ref{Fig-3-Box-Summary-DAGs}\,\textit{(d)} reconstructing, via Eq.~(\ref{Joint-Distr-D}), the full statistics in Fig.~\ref{Fig-3-Box-Behaviour} for all $C=1,2,3$.
\end{proof}

\end{document}